\documentstyle{mn}
\input{psfig}
\newif\ifAMStwofonts

\def\simgt{\hbox{\rlap{\raise 0.425ex\hbox{$>$}}\lower 0.65ex\hbox{$\sim$}}}
\def\simlt{\hbox{\rlap{\raise 0.425ex\hbox{$<$}}\lower 0.65ex\hbox{$\sim$}}}

\def\degree{^\circ}

\def\bj {b_{\rm J}}

\def\civ {C{\small IV}}

\ifoldfss
  \ifCUPmtlplainloaded \else
    \NewTextAlphabet{textbfit} {cmbxti10} {}
    \NewTextAlphabet{textbfss} {cmssbx10} {}
    \NewMathAlphabet{mathbfit} {cmbxti10} {} 
    \NewMathAlphabet{mathbfss} {cmssbx10} {} 
  \fi
  \ifAMStwofonts
    \ifCUPmtlplainloaded \else
      \NewSymbolFont{upmath} {eurm10}
      \NewSymbolFont{AMSa} {msam10}
      \NewMathSymbol{\upi}     {0}{upmath}{19}
      \NewMathSymbol{\umu}     {0}{upmath}{16}
      \NewMathSymbol{\upartial}{0}{upmath}{40}
      \NewMathSymbol{\leqslant}{3}{AMSa}{36}
      \NewMathSymbol{\geqslant}{3}{AMSa}{3E}

    \fi
  \fi
\fi 

\ifnfssone
  \newmathalphabet{\mathit}
  \addtoversion{normal}{\mathit}{cmr}{m}{it}
  \addtoversion{bold}{\mathit}{cmr}{bx}{it}
  \newmathalphabet{\mathbfit} 
  \addtoversion{normal}{\mathbfit}{cmr}{bx}{it}
  \addtoversion{bold}{\mathbfit}{cmr}{bx}{it}
  \newmathalphabet{\mathbfss} 
  \addtoversion{normal}{\mathbfss}{cmss}{bx}{n}
  \addtoversion{bold}{\mathbfss}{cmss}{bx}{n}
  \ifAMStwofonts
    \ifCUPmtlplainloaded \else
      \UseAMStwoboldmath
      \makeatletter
      \new@mathgroup\upmath@group
      \define@mathgroup\mv@normal\upmath@group{eur}{m}{n}
      \define@mathgroup\mv@bold\upmath@group{eur}{b}{n}
      \edef\UPM{\hexnumber\upmath@group}
      \new@mathgroup\amsa@group
      \define@mathgroup\mv@normal\amsa@group{msa}{m}{n}
      \define@mathgroup\mv@bold\amsa@group{msa}{m}{n}
      \edef\AMSa{\hexnumber\amsa@group}
      \makeatother
      \mathchardef\upi="0\UPM19
      \mathchardef\umu="0\UPM16
      \mathchardef\upartial="0\UPM40
      \mathchardef\leqslant="3\AMSa36
      \mathchardef\geqslant="3\AMSa3E
    \fi
  \fi
\fi 

\ifnfsstwo
  \DeclareMathAlphabet{\mathbfit}{OT1}{cmr}{bx}{it}
  \SetMathAlphabet\mathbfit{bold}{OT1}{cmr}{bx}{it}
  \DeclareMathAlphabet{\mathbfss}{OT1}{cmss}{bx}{n}
  \SetMathAlphabet\mathbfss{bold}{OT1}{cmss}{bx}{n}
  \ifAMStwofonts
    \ifCUPmtlplainloaded \else
      \DeclareSymbolFont{UPM}{U}{eur}{m}{n}
      \SetSymbolFont{UPM}{bold}{U}{eur}{b}{n}
      \DeclareSymbolFont{AMSa}{U}{msa}{m}{n}
      \DeclareMathSymbol{\upi}{0}{UPM}{"19}
      \DeclareMathSymbol{\umu}{0}{UPM}{"16}
      \DeclareMathSymbol{\upartial}{0}{UPM}{"40}
      \DeclareMathSymbol{\leqslant}{3}{AMSa}{"36}
      \DeclareMathSymbol{\geqslant}{3}{AMSa}{"3E}
    \fi
  \fi
\fi 

\ifCUPmtlplainloaded \else
  \ifAMStwofonts \else 
    \def\upi{\pi}
    \def\umu{\mu}
    \def\upartial{\partial}
  \fi
\fi

\title[The 2QZ 10k catalogue]{The 2dF QSO Redshift Survey - V.  The 10k
catalogue}

\author[S.M.Croom et al.]
	{S.M. Croom$^1$\thanks{E-mail: scroom@aaoepp.aao.gov.au},
	R.J. Smith$^2$, B.J. Boyle$^1$, T. Shanks$^3$, N.S. Loaring$^4$, L. Miller$^4$,\\
	{\LARGE I.J. Lewis$^1$}\\  
${^1}$ Anglo-Australian Observatory, PO Box 296, Epping, NSW 1710, 
Australia \\ 
${^2}$ Liverpool John Moores University, Twelve Quays House, Egerton Wharf,
Birkenhead, CH41 1LD, UK\\
${^3}$Department of Physics, University of Durham, South Road, 
Durham, DH1 3LE\\
${^4}$ Department of Physics, Oxford University, 1 Keble Road, Oxford,
OX1 3RH}

\begin{document}

\maketitle

\newcommand{\fmmm}[1]{\mbox{$#1$}}
\newcommand{\scnd}{\mbox{\fmmm{''}\hskip-0.3em .}}
\newcommand{\scnp}{\mbox{\fmmm{''}}}

\begin{abstract}

We present a catalogue comprising over 10000 QSOs covering an
effective area of 289.6 deg$^2$, based on spectroscopic observations
with the 2-degree Field instrument at the Anglo-Australian Telescope.
This catalogue forms the first release of the 2-degree Field QSO
Redshift Survey.  QSO candidates with $18.25 < \bj < 20.85$ were
obtained from a single homogeneous colour-selected catalogue based on
APM measurements of UK Schmidt photographic material.  The final
catalogue will contain approximately 25000 QSOs and will be released
to the public at the end of 2002, one year after the observational
phase is concluded.
\end{abstract}
\section{Introduction}

The aim of the 2dF QSO Redshift Survey (2QZ) is to measure redshifts
for $\sim 25000$ optically-selected QSOs with $\bj<20.85$ and $z<3$.
The spectra have been obtained with the 2-degree Field (2dF)
instrument (Lewis, Glazebrook \& Taylor 1998)\nocite{lgt98} on the
Anglo-Australian Telescope (AAT).  The  final catalogue will represent
a factor $\sim50$ increase in the number of  QSOs identified in a single
homogeneous survey to this magnitude limit (c.f. The Durham/AAT
survey; Boyle et al. 1990).  The primary science goal of this survey
is to obtain an accurate measure of the large-scale structure in the
Universe out to high redshifts, $z<3$, and large scales,
$<1000$h$^{-1}$Mpc, via the study of QSO clustering.

As of January 2001, the survey is over 60 per cent complete. To
expedite community access to the 2QZ, we are releasing the catalogue
in two stages.  This paper describes the first release containing
20590 objects, including 11005 QSOs, for which we have obtained 2dF
spectra. The release represents the `most complete' subset of the
entire survey observed prior to December 2000 -- containing all fields
for which we have achieved a spectroscopic completeness of 85 per cent
or greater.  This catalogue has been used to derive results on the QSO
correlation function (Croom et al.\ 2001, Paper II), the power
spectrum (Hoyle et al.\ 2001, Paper IV) and the composite spectrum
(Smith et al.\ 2001, Paper VI, in preparation).  Results on the QSO
luminosity function from the 2QZ (Boyle et al.\ 2000, Paper I) were
based on an earlier version of the catalogue comprising $\sim 5000$
QSOs.  When complete the catalogue will contain approximately 48000
objects, including 25000 QSOs.  The final catalogue will be released
at the end of 2002, one year after the observational phase of the
survey is concluded, currently planned for the end of 2001.

This paper describes the first release catalogue, hereinafter the 2QZ
10k catalogue, of the 2QZ.  Sections 2 and 3 describe the
spectroscopic observations and catalogue format.  Section 4 then
discusses estimates of the catalogue quality, and presents a new
estimate of the QSO luminosity function.
 
\section{Spectroscopic Observations}

\subsection{The candidate catalogue}

QSO candidates with $18.25 < \bj < 20.85$ were selected for
observation with 2dF from measurements of $u\bj r$ UK Schmidt
photographic material.  Full details are given in Smith et al.\ (2001,
Paper III).  The 2QZ area comprises 30 UKST fields, arranged in two
$75\degree \times 5\degree$ declination strips centred on $\delta =
-30\degree$ and $\delta = 0\degree$.  The $\delta = -30\degree$ strip
extends from $\alpha$ = 21$^{h}$40 to $\alpha$ = 3$^{h}$15 in the
South Galactic Cap and the equatorial strip from $\alpha$ = 9$^{h}$50
to $\alpha$ = 14$^{h}$50 in the North Galactic Cap. When complete, the
total survey area will be 740 deg$^{2}$ allowing for regions of sky
excised around bright stars.  The 2QZ area forms an exact subset of
the 2dF Galaxy Redshift Survey (2dFGRS; see Colless 1998) area. 2dF
observations of the 2QZ were therefore combined with those  of the
2dFGRS.  A tiling algorithm devised by the 2dFGRS team was used to
determine the positions of individual 2dF field centres in order to
obtain near-complete spectroscopic coverage (95 per cent for galaxies,
99 per cent for QSO candidates) for the 2QZ and 2dFGRS targets in the
survey area with the minimum number of individual 2dF pointings.

The 2QZ colour selection is designed to be largely complete ($>90$ per
cent) for QSOs with $0.3<z<2.2$.  At higher redshifts, $2.2<z<3.0$,
the photometric completeness of the survey gradually drops, with the
highest redshift QSOs in the 2QZ being found at $z\simeq3$.
Paper I  provides an estimate of the completeness contours based on
$u\bj r$ selection as a function of $\bj$ and $z$.  The stellar
morphological criterion also causes incompleteness at low redshifts.
This is more difficult to estimate but it is likely to cause some form
of incompleteness up to $z\sim 0.5$.  The photometric errors in the
magnitudes have already been discussed extensively in Paper III.  They
are approximately constant over the  magnitude range of the survey
$\sigma(\bj)=\sigma(r)=0.1$, $\sigma(u)=0.15$.  Independent CCD
sequences have been used to calibrate each UKST field (Boyle, Shanks
\& Croom 1995; Croom et al. 1999).

\subsection{2dF observations}

Observations of the 2QZ began in October 1997.  This initial release
is based on 304 individual 2dF fields observed over the period October
1997 to November 2000.  The two 2dF spectrographs each contain one
1024$\times$1024 Tektronix CCD with $24\mu$m pixels and are each fed
by 200 optical fibres.  2QZ objects were observed with 2dF using the
low resolution 300B gratings, providing a dispersion of
178.8\AA~mm$^{-1}$ (4.3\AA~pixel$^{-1}$) and a resolution of
$\simeq8.6$\AA\ ($\simeq2$ pixels FWHM) over the range 3700\AA --
7900\AA.  Each optical fibre has a diameter on the sky of $2.16''$ at
the centre of a field, falling to $2''$ at the edge of the field.  The
fibre spectra are projected to $\sim1.5$ pixels (FWHM) on the detector
and are each separated by $\sim5$ pixels.  Each field received an
approximately 55 minute integration (3 $\times$ 1100$\,$sec), combined
with a flat-field, used to determine the positions of the fibre
spectra on each CCD frame, and a CuArHe arc for wavelength
calibration.  The fields were observed in a wide variety of seeing
conditions (1 arcsec -- 2.5 arcsec FWHM) and transparency (details of
conditions are included in the data release).  Where possible,
integration times were adjusted to match conditions.  However, the
instrumental requirement to observe 2dF fields at the hour angle for
which they had been configured limited the opportunity to extend
integration times significantly.  As a result the quality of data can
vary significantly from field to field.  We have attempted to minimize
the effects of this variation in the 10k release by only including
objects from fields where at least 85 per cent of the objects observed
yielded a spectroscopic identification.

Data from 2dF was reduced using the pipeline reduction system {\small
2DFDR} \cite{2dfman}.  CCD frames from each spectrograph
were reduced separately.  Following bias removal, fibre extraction and
wavelength calibration, the transmissions of each fibre
were normalized using the relative strengths of the night sky
lines in each fibre spectrum.  Sky subtraction  was then done by
subtracting a median sky spectrum determined from up to 20 sky-dedicated
fibres in each frame (the minimum number of sky fibres used was six
per spectrograph).

\section{The catalogue}

\subsection{Spectroscopic identification}

\begin{figure*}
\centering
\centerline{\psfig{file=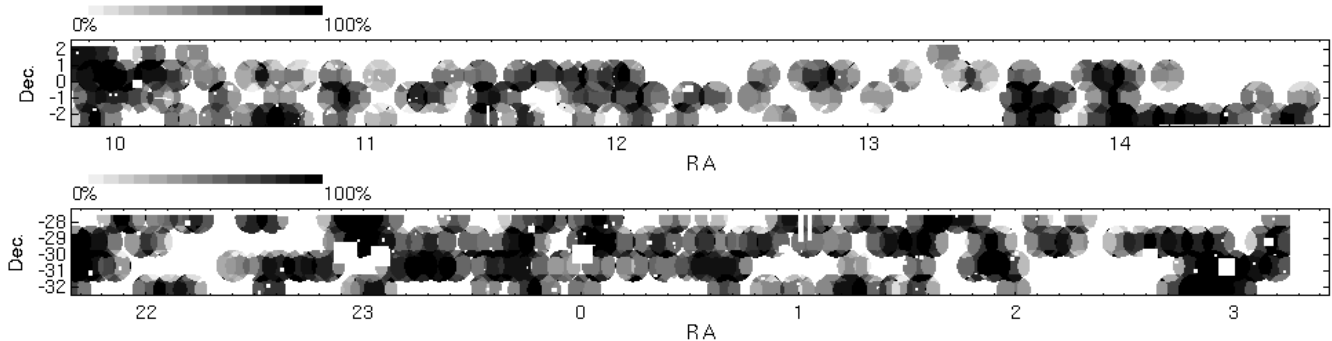,angle=270,width=18cm}}
\caption{The coverage map for the 2QZ 10k catalogue.  The grey-scale indicates 
the observational completeness as a function position on the sky.  The tiling
pattern of 2dF pointings used to cover the survey area can be seen clearly
from this figure.}
\label{fig:coverage}
\end{figure*}

Reduced spectra of QSO candidates were then passed through a
non-interactive program {\small AUTOZ} (Miller et al. in preparation)
to classify individual objects and measure a redshift for those that
turned out to be QSOs or galaxies.  The $u\bj r$ colour-selection
process used to generate the input catalogue has non-negligible
contamination (originally estimated to be $\sim 45$ per cent, see
Paper III) from galactic stars (subdwarfs and white dwarfs) and
compact emission line galaxies.  {\small AUTOZ} fits a variety of
templates to each observed spectrum.  These templates include QSOs,
both `normal' and broad absorption line (BAL) QSOs, emission line
galaxies and a library of stellar spectra including main sequence
types from O to M and DA, DB and DZ white dwarfs.  The {\small AUTOZ}
program uses the fact that data are collected on a large number of
different objects (QSOs, galaxies and stars) at once ($\sim200$ per
spectrograph) to estimate the wavelength dependent sensitivity and
noise.  These terms are then taken into account when carrying out a
$\chi^2$ fit to the templates to derive the best fit model.

In general {\small AUTOZ} produces reliable identifications for a wide
variety of the objects.  However, we have carried out manual checks of
all spectra in the survey to identify the small fraction of objects
with peculiar or low signal-to-noise (S/N) spectra which {\small
AUTOZ} had difficulty in identifying correctly.  Each member of the
2QZ team independently examined 5000--6000 spectra by eye and corrected any
{\small AUTOZ} identifications that were clearly in error.  As a
further double-check, two members of the survey (BJB and SMC)
re-examined all spectra  to ensure a degree of homogeneity in this
re-classification procedure.

The sources are classified into six categories:

\begin{tabbing}
12345678\=\kill

{\bf QSO} \> Spectrum with one or more broad ($>1000\,$km$\,$s$^{-1}$)\\
\>emission lines.\\   
{\bf NELG} \> Galaxy spectrum with one or more narrow \\ 
\>($<1000\,$km$\,$s$^{-1}$) emission lines.\\
{\bf gal} \> Galaxy spectrum with no emission lines.\\
{\bf star} \> Galactic star spectrum.\\
{\bf cont} \> High S/N spectrum (S/N$>10$) with no\\ 
\> identifiable emission or absorption features.\\
{\bf ??}\>Unclassifiable spectrum due to low S/N.\\
\end{tabbing}

The quality of the identification and redshift estimation is divided
into three levels of confidence: 
\begin{itemize}
\item Quality 1: High quality identification or redshift
\item Quality 2: Low quality identification or redshift
\item Quality 3: No classification or redshift assignment
\end{itemize}
A quality flag is assigned independently to both the identification
and redshift of an object.  Approximately 6 per cent of objects were
found to have low quality identifications (quality 2), these objects
appear in the final catalogue with a `?' against their identification
(apart from objects identified as cont, which all have quality 2
identifications).

\begin{figure}
\centering
\centerline{\psfig{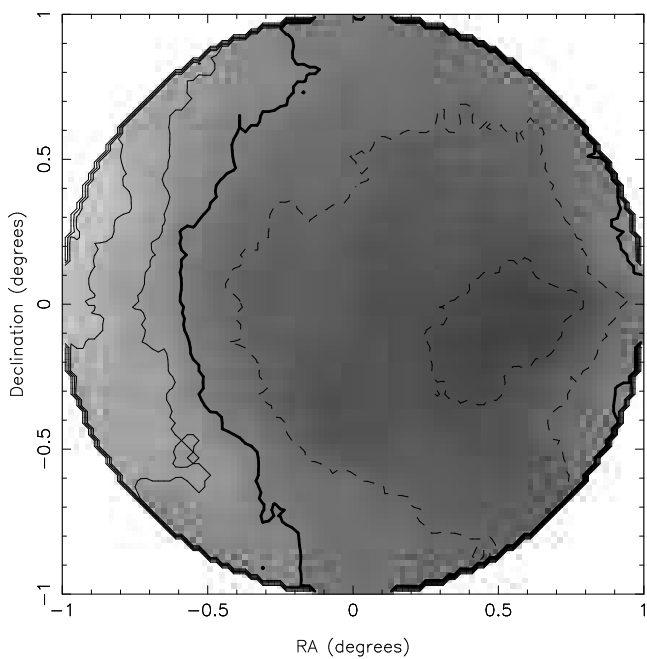}}
\caption{The distribution of spectroscopic magnitude minus catalogue
magnitude across a 2dF field for sources in the 10k catalogue.
Contours are at $\Delta{\rm mag}=0.2$ intervals, the thick lined
contour being at $\Delta{\rm mag}=0$ and the dashed line contours
being at $\Delta{\rm mag}<0$.}
\label{fig:deltamag}
\end{figure}

For each 2dF pointing we calculate a {\it spectroscopic} completeness,
being the ratio of objects observed in the field with quality 1
or quality 2 identifications to the total number of spectroscopically
observed objects.  Stars are likely to dominate the
unidentified population, at least for low levels of incompleteness.
This is because their weak spectral features make them very difficult
to identify from low S/N spectra.  In contrast, QSOs with broad
emission  lines should be amongst the easiest class of objects to
identify from low S/N spectra.   To establish the spectroscopic
incompleteness level at which we also begin to lose significant
numbers of QSOs, we compared the normalized QSO number-magnitude
relation obtained from fields with greater than 95 per cent
completeness to those from fields with completenesses in the range
90 -- 95 per cent, 85 -- 90 per cent, 80 -- 85 per cent and 75 -- 80
per cent.  Using a Kolmogorov-Smirnov test, we found that the
$N(m)$ showed a significant difference (at the 99 per cent confidence
level) for fields with lower than 85 per cent completeness.

We therefore chose to include only those fields with spectroscopic
completenesses of 85 per cent or greater in this initial release.
This corresponds to 233 2dF fields.  The coverage provided by these
fields is indicated in Fig. \ref{fig:coverage} (taken from Paper II).
Note that many areas of sky have not yet received their full
complement of 2dF pointings in the adaptive tiling strategy.  As a
result, the current survey exhibits a spatially-dependent
observational incompleteness.  Hence, although these fields have 85
per cent or greater identification rate, there remains a still larger
fraction of incompleteness due to sources which have yet to be
observed.  The 2QZ coverage map shown in Fig. \ref{fig:coverage}
(released in digital form as a survey product) can be used to account for this.  This
coverage map is produced by calculating the ratio of spectroscopically
observed objects to input catalogue objects in each region defined by
the intersection of 2dF pointings.  This distribution is then
pixelized using $1\times1$ arcminute pixels to produce a final
observational completeness map.

We also check whether there is any variation in the catalogue over the
area of one 2dF field.  In Fig. \ref{fig:deltamag} we plot the mean
difference ($\Delta$mag) between the catalogue $\bj$ magnitude and a
$\bj$-band magnitude estimated from the spectroscopic data averaged
over all the 2dF fields in the 10k catalogue (note that the averaging is
carried out in flux, not magnitude).  This is only a relative
difference as the zero-points of the spectroscopic magnitudes are
derived from the mean differences in each 2dF field.  A positive
$\Delta$mag indicates a spectroscopic magnitude which is fainter than
the catalogue magnitude.  We see that there is deficit of flux at the
edge of the field (particularly on the left side).  This is probably
due to residual errors that has not been completely accounted for
by the 2dF astrometric model.  This should not significantly affect
the spatial distribution of QSOs in our sample for two reasons.
First, the tiling algorithm used to place the fields produces
significant overlap so that any effect in a 2dF field area will be
blurred out.  Second, if we measure the identification rate as a
function of field position, the only significant variation is at
radii $>0.7^{\circ}$.  On scales smaller than this the mean
completeness is $\simeq95$ per cent.  At the extreme edges of the
field, radii $>0.9^{\circ}$, the mean completeness only drops to
$\simeq87$ per cent.  This variation in completeness, internal to 2dF
fields is of the same order as the variations in completeness between
2dF fields.

\subsection{10k catalogue description}

\begin{table}
\centering
\caption{Composition of the 2QZ catalogue}
\label{tab:class}
\begin{tabular}{@{}lrrrr@{}}
\hline
Identification&Number&Quality 1&Quality 2&Percentage\\
\hline
QSO         & 11005 & 10689 & 316  & 53.4\\
NELG        & 2104  & 2059  & 45   & 10.2\\
star        & 5842  & 5081  & 761  & 28.4\\
gal         & 44    & 37    & 7    & 0.2\\
cont        & 108   & --    & 108  & 0.5\\
??          & 1487  & --    & --   & 7.2\\
{\bf Total} & 20590 & 17866 & 1237 &\\
\hline
\end{tabular}
\end{table}

\begin{figure*}
\centering
\centerline{\psfig{file=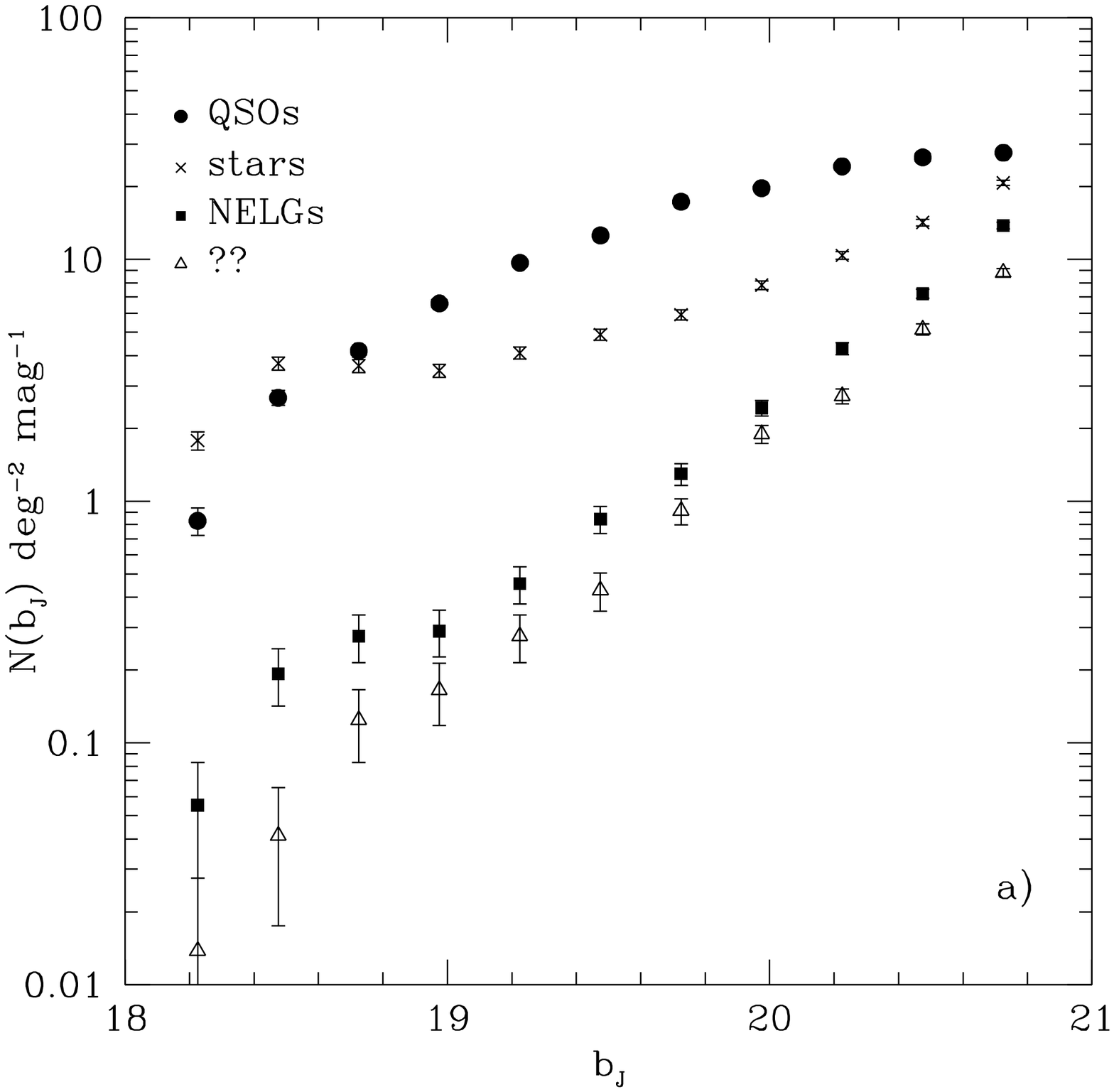,width=8cm}\psfig{file=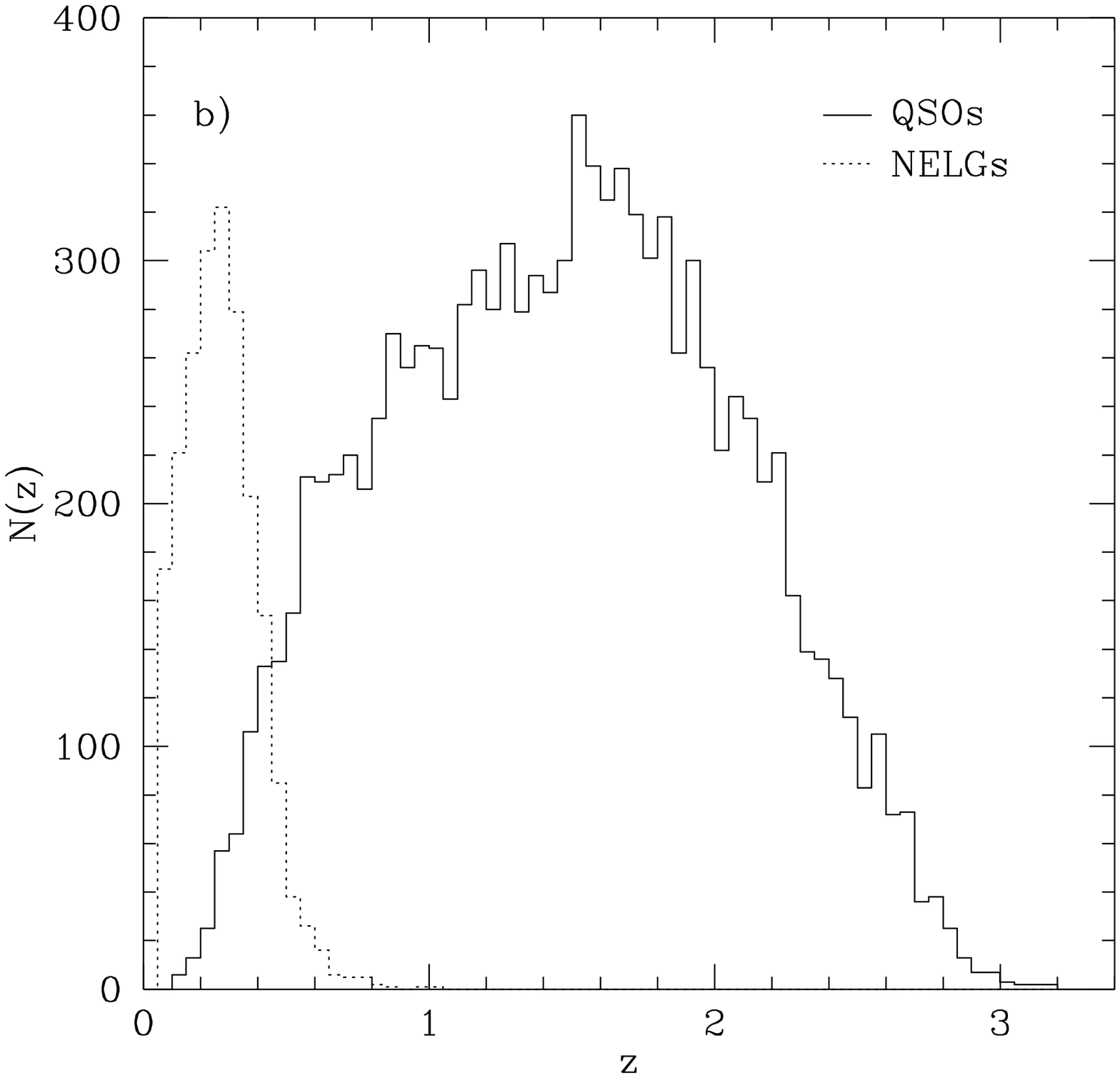,width=8cm}}
\caption{The number-magnitude and number-redshift distributions for
the sources in the 10k catalogue.  a) The number-magnitude
distribution for QSOs (filled circles), stars (crosses), NELGs
(filled squares) and unidentified objects (open triangles).  b) The
number-redshift distribution for QSOs (solid line) and NELGs (dotted
line).  No corrections for spectroscopic or photometric incompleteness
are made in either figure.}
\label{fig:nz}
\end{figure*}

Based on the criteria defined above, the initial release of the 2QZ
catalogue contains entries for 20590 objects, including 11005 QSOs.
The composition of the 10k catalogue as a function of spectroscopic
class is given in Table~\ref{tab:class}.  As defined by the fraction
of objects with quality 1 or 2 classifications, the mean overall
spectroscopic completeness of the catalogue is 93 per cent.  The
effective area of the survey (after correction for the observational
incompleteness) is 289.6 deg$^2$.  The distribution of sources as a
function of $\bj$ magnitude is shown in Fig. \ref{fig:nz}a.  QSOs
dominate the source counts at all but the  brightest magnitudes.  The
QSOs show a smooth redshift distribution over the full range of the
survey, $0.1<z<3.0$, with a maximum at the 2QZ median redshift
$z\sim 1.5$ (Fig. \ref{fig:nz}b).

The released catalogue is available as an ASCII file from  {\tt
http://www.2dfquasar.org}.  The catalogue format is given in
Table~\ref{tab:cat}.  The first part of a catalogue entry contains
details from the input catalogue, such as position and magnitude.  We
note in particular that sources which had only upper limits
(i.e. non-detections) on the $r$ plates are also included in the
catalogue and have a listed $\bj-r$ colour.  In this case the colour
term is $(\bj-r_{\rm lim})-10$, the $-10$ being used to differentiate
upper limits from normal colours (objects with real $r$-band
detections have colours in the range $-1.4<\bj-r<3.4$, while upper
limits have $\bj-r<-9.8$).  The input catalogue information is followed by
details of observations and identifications.  In a number of cases
($\sim1200$ objects, $\sim5\%$ of the catalogue) we have two
observations for a source.  These are useful to assess the quality of
the final catalogue (see below).  The identification and redshift that
we adopt is that with the lowest quality value (where the quality
value is identification quality$\times10$ + redshift quality), and if
there are equal quality values, the highest S/N.  In all cases the
final adopted ID is listed as observation \#1, with the lower quality
observation being listed as observation \#2.  Finally we also include
cross-matches to selected other databases.  Where a 2QZ source matches
the position (to within $6''$) of a previously known QSO/AGN in the
catalogue of V\'eron-Cetty \& V\'eron (2000) we include the previously
known redshift.  We also include radio fluxes at 1.4GHz from the NRAO
VLA Sky Survey (NVSS; Condon et al. 1998) and X-ray fluxes from the
ROSAT All Sky Survey (RASS; Voges et al. 1999; Voges et al. 2000),
converting from RASS counts per second to flux (in
erg$\,$s$^{-1}$cm$^{-2}$) by multiplying by $5.6\times10^{-12}$.
Lastly we include an estimate of the galactic reddening $E(B-V)$ to
each source taken from the work of Schlegel, Finkbeiner \& Davis (1998).

\begin{table*}
\centering
\caption{Format for the 2QZ catalogue.  The format entries are based
on the standard {\small FORTRAN} format descriptors.}
\label{tab:cat}
\begin{tabular}{@{}lrl@{}}
\hline
Field&Format&Description\\
\hline
Name     &          a20 & IAU format object name\\
RA       &   i2 i2 f5.2 & RA J2000 (hh mm ss.ss)\\
Dec      & a1i2 i2 f4.1 & Dec J2000 ($\pm$dd mm ss.s)\\
RA       &         f8.6 & RA J2000 (radians) \\
Dec      &         f9.6 & Dec J2000(radians) \\
$\bj$    &         f5.2 & $\bj$ magnitude\\
$u-\bj$  &         f5.2 & $u-\bj$ colour\\
$\bj-r$  &         f6.2 & $\bj-r$ colour [including $r$ upper limits
as: $(\bj-r_{\rm lim})-10.0$]\\
$N_{obs}$&           i1 & Number of observations\\
\hline
Observation \# 1\\
$z_1$    &         f5.3 & Redshift\\
q$_{1}$  &           i2 & Identification quality $\times$ 10 + redshift quality\\
ID$_1$   &          a11 & Identification\\
date$_1$ &           a8 & Observation date\\
fld$_1$  &           i4 & 2dF field number $\times 10$ + spectrograph number\\
$S/N_1$  &         f6.2 & Signal-to-noise ratio in 4000--5000{\AA} band\\
\hline
Observation \# 2\\
$z_2$    &         f5.3 & Redshift\\
q$_{2}$  &           i2 & Identification quality $\times$ 10 + redshift quality\\
ID$_2$   &          a11 & Identification\\
date$_2$ &           a8 & Observation date\\
fld$_2$  &           i4 & 2dF field number $\times 10$ + spectrograph number\\
$S/N_2$  &         f6.2 & Signal-to-noise ratio in 4000--5000{\AA} band\\
\hline
$z_{\rm prev}$&    f5.3 & Previously known redshift (V\'eron-Cetty \& V\'eron 2000)\\
radio    &         f6.1 & 1.4GHz Radio flux, mJy (NVSS) \\
x-ray    &         f7.4 & X-ray flux, $\times10^{-13}\,$erg$\,$s$^{-1}$cm$^{-2}$ (RASS)\\
dust     &         f5.3 & $E(B-V)$ (Schlegel et al. 1998)\\
\hline
\end{tabular}
\end{table*}

We do not distinguish between different narrow-emission-line galaxy
classifications (e.g. LINER, Seyferts) in the catalogue.  Also, the
spectral coverage provided by 2dF limits our ability to detect
broad MgII $\lambda2798$ emission until $z>0.35$.  In some QSOs with
$0.35<z<0.5$, broad MgII is clearly detected with no corresponding
broad H$\beta$ seen.  Thus at lower redshifts, $0.15<z<0.35$, some
objects classified as NELGs on the basis of a narrow H$\beta$ line may
exhibit broad MgII below the blue limit of the 2dF spectral window.
For QSOs with $z>1.6$, {\small AUTOZ} fits both `normal' and BAL QSO
templates.  Objects identified as BALs are indicated by `QSO(BAL)' in
the identification column.  However, due to the varied nature of BAL
QSOs, this will not be a comprehensive list of BALs.  A more detailed
search for BALs has not yet been applied.

Within the objects we classify as stars, in the main we only give
classifications for white dwarfs (WDs).  We do not attempt spectral
classification of main sequence stars.  The $\sim1000$ WDs are mostly
DAs with strong broad hydrogen Balmer absorption lines.  At low
S/N these can be confused with A stars.  Only detailed
fitting of the temperature and gravity of these stars can easily
resolve this issue, and this is beyond the scope of the present
paper.  As well as DAs we also find a number of DB and DO WDs which
are dominated by neutral and singly ionized helium respectively.  A
small number of DZs are also found, with broad calcium H and K
absorption.  The only other stellar types we classify are 6 emission
line stars with strong hydrogen Balmer emission which we classify as
CV (cataclysmic variable) and WD + M dwarf binary systems which are
denoted as DA/M.  We note that our code has been designed primarily to
simply identify an object as a star (and therefore not a QSO), and
that a more detailed analysis would no doubt provide more accurate
classifications.  The above classifications are denoted within the
catalogue in parentheses after the main star identification
[e.g. star(DB), star(CV)].

\begin{figure*}
\centering
\centerline{\psfig{file=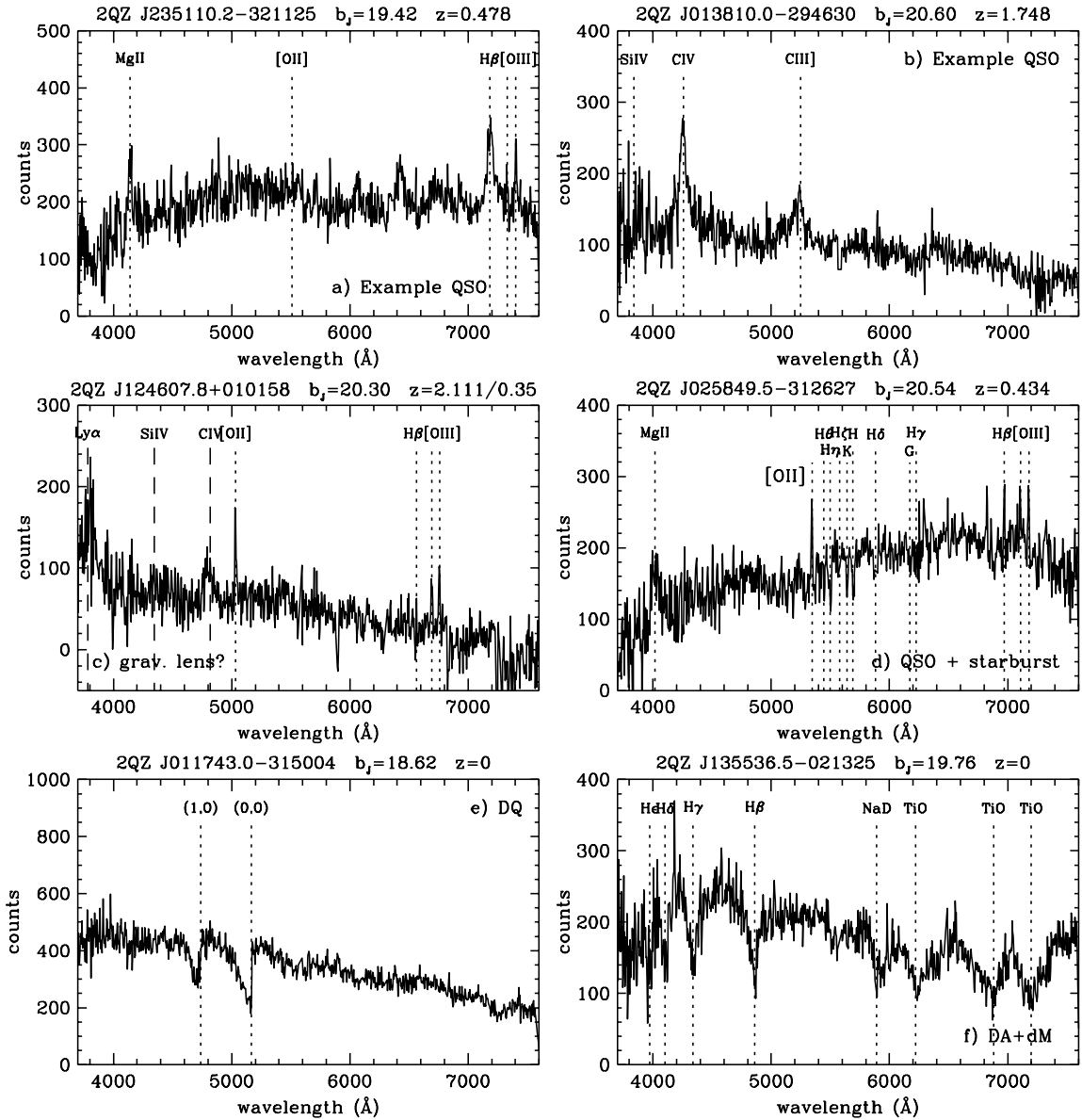,width=18cm}}
\caption{ A selection of spectra taken from the 2QZ 10k catalogue.  a)
and b) two randomly selected example QSO spectra, c) a potential
gravitational lens, with QSO  emission lines at $z=2.111$ (dashed
lines) and NELG emission lines at $z=0.350$ (dotted lines), d) a QSO +
starburst composite, e) a DQ white dwarf with the molecular carbon
(Swan) bands indicated, f)  a white dwarf M-dwarf binary, showing
broad hydrogen Balmer absorption in the blue and TiO absorption bands
in the red.}
\label{fig:qsospec}
\end{figure*}

To allow others to carry out more detailed spectral analysis of both
QSOs and other sources, spectra for all objects in the 2QZ 10k
catalogue are provided as a primary data product in this release.
This includes objects for which no identification could be made.  Many
of these are low S/N spectra, but objects also remained unclassified
for a number of other reasons.  Occasionally a damaged fibre will
produce fringing which shows up as a strong oscillation as a function
of wavelength, rendering the underlying spectrum unusable.  Also
badly subtracted night sky emission (and in some cases moonlight) can
sometimes seriously affect our ability to identify an object.  The
spectra are available in FITS format from the same location as the
catalogue.  Fig.  \ref{fig:qsospec} shows some example 
spectra from the 10k catalogue.  Figs. \ref{fig:qsospec}a and b are
randomly selected QSO spectra. Figs. \ref{fig:qsospec}c-f show a
range of peculiar spectra.  Fig. \ref{fig:qsospec}c shows
narrow emission lines implying a redshift of $z=0.350$ and broad
emission lines suggesting a redshift of $z=2.111$.  This appears to
be a superposition of a QSO and a NELG, and as such is a potential
gravitational lens.  Fig. \ref{fig:qsospec}d shows the spectrum of a
starburst + QSO, a less extreme version of the post-starburst object
found in the cross-matching of the 2QZ and the NVSS by Brotherton et
al. (1999).  Fig. \ref{fig:qsospec}e is a rare carbon (DQ) white dwarf
and Fig. \ref{fig:qsospec}f is a white dwarf-M dwarf binary system.

\section{Catalogue Quality}

\subsection{Identification}

The overlap between the 2dF pointing has resulted in a number of
objects in the 2QZ catalogue for which we have two independent 2dF
observations.  By comparing the identification/redshift estimate for
different observations we can obtain an estimate of the reliability  of
the identification and accuracy of the redshift determination.  For
the 1019 objects with quality 1 or 2 identifications and more than one
2dF observations in fields included in the 10k catalogue, we find 87
whose identifications differ between the observations (excluding the 5
objects whose identification changed from QSO to NELG but with the
same redshift).  This corresponds to a mis-identification rate of 8.5
per cent.  If we restrict our attention to those objects with quality
1 identifications for both observations, this mis-identification rate
drops to 3.8 per cent (34 objects with different classifications out
of 889 objects).

The mis-identification rate amongst class 2 identifications is higher.
If we directly compare the identifications for those objects with a
quality 2 identification from both observations, we find 8 objects
with different identifications from a total sample of 22 objects;
equivalent to a mis-classification rate of 36 per cent.  Comparison
between quality 1 observations and quality 2 observations gives
similar results.  Of the 108 objects given an identification quality
of 1 in one observation and a quality of 2 in the other, 45 (42 per
cent) changed their identification in the subsequent higher quality
observation.  Thus the reliability of quality 2 identifications is
approximately 60 per cent.  
 
\begin{figure}
\centering
\centerline{\psfig{file=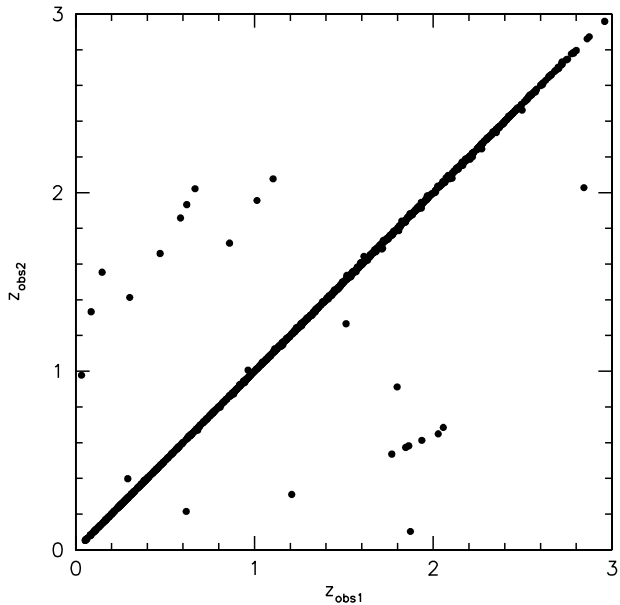,width=8.5cm}}
\caption{Comparison between redshifts obtained for different 
observations of the same object.}
\label{fig:diffz}
\end{figure}

A more quantitative measure of the quality of the spectrum (although not
necessarily the reliability of the identification) is the
signal-to-noise.  If we restrict our comparison to objects where both
spectra have a S/N$>5$ per pixel, we find a overall
mis-classification rate of 1.7 per cent (6 objects out of 359).
 
\subsection{Redshift accuracy}

We can also obtain an estimate of the reliability of the redshift
determination from these repeat observations.  For all
quality 1 and 2 QSO, NELG or galaxy identifications there are
21 cases (out of 602 objects) where the redshift estimate 
differs by more than 5 per cent between the different observations.
The comparison between the redshifts obtained from the different observations 
is shown in Figure~\ref{fig:diffz}.  All objects with significantly different
redshifts are classified as QSOs; in most cases the difference in the 
redshift estimate is caused by the incorrect identification of a single
emission line seen in the QSO spectrum.  In total there are
518 QSOs with two observations, thus the reliability rate for
QSO redshifts is 97 per cent and effectively 100 per cent for
NELGs and galaxies.   

Excluding objects with incorrect redshifts due to the mis-identification
of emission lines, the rms difference between the redshift measurements
is $0.005$.  

We also compared our identifications with those objects which had been
previously classified as QSOs.  Based on positional matches with QSOs
in the V\'eron-Cetty \& V\'eron (2000) catalogue, we found that 331
objects in the 2QZ 10k catalogue had been previously identified as
QSOs.  These objects are flagged in the released catalogue (see
Table~\ref{tab:cat}).  Of these, 305 were confirmed as QSOs by 2QZ.
However, 33 QSOs were found to have redshifts which disagreed
significantly between the 2QZ and previous estimates.  Visual
inspection of the 2QZ spectrum revealed that, in all but one case, the
2QZ redshift was secure (i.e. two or more strong emission lines
correctly identified).  The remaining 29 objects were not classified
as QSOs in the 2QZ, these identifications included 7 stars (2 white
dwarfs), 4 star?s, 3 NELGs, 1 cont, 1 cont?, 1 QSO? and 13 with
unclassifiable spectra.  Visual inspection of the quality 1 spectra
confirmed the 2dF classification in all cases. 

In summary, the comparison between independent observations of over
1000 objects in the 2QZ release (approximately 5 per cent of the
catalogue) reveals the overall reliability of the identifications
is over 90 per cent.  For quality 1 classifications this figure
is close to 97 per cent; for quality 2 classifications the reliability
is approximately 60 per cent.  The redshift reliability for QSOs is 97
per cent with an rms error in the redshift measurement of $\sigma(z) =
0.005$.

\begin{figure}
\centering
\centerline{\psfig{file=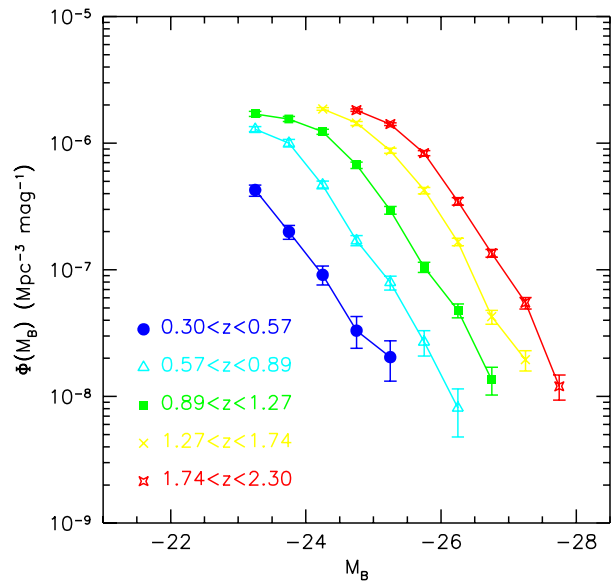,width=8.5cm}}
\caption{The optical QSO luminosity function derived from the 10k
catalogue and the LBQS for an Einstein-de Sitter universe. Only QSOs
with $M_B<-23$ and $0.3<z<2.3$ are included in this analysis.}
\label{fig:lf}
\end{figure}

\subsection{The 10k luminosity function}

For completeness, we have also repeated the analysis of Paper I to
derive a luminosity function (LF) of the QSOs in the 10k catalogue,
including objects  from the Large Bright Quasar Survey (LBQS; Hewett,
Foltz \& Chaffee 1995) at  bright magnitudes. In total, 9633 QSOs with
$M_B<-23$ and $0.3<z<2.3$ (8743 from the 2QZ and 890 from the LBQS)
were used to obtain the binned LF  for an Einstein-de Sitter universe
($H_0=50\,$km$\,$s$^{-1}\,$Mpc$^{-1}$) which is shown in Fig. \ref{fig:lf}.  

\begin{figure}
\centering
\centerline{\psfig{file=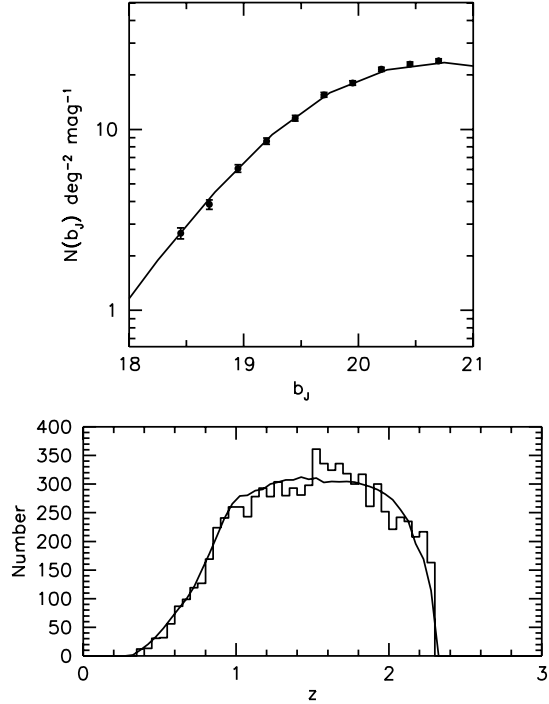,width=10.0cm}}
\caption{Upper panel: derived number-magnitude, $N(m)$, relation
for QSOs with $M_B<-23$ and $0.3<z<2.3$ in the 2QZ survey (filled
dots) and the prediction from the best fit polynomial luminosity
evolution model.  The data has been corrected for photometric and
spectroscopic incompleteness as in Paper I.  Lower panel: observed
number-redshift relation, $N(z)$, for the 2QZ QSOs with $M_{\rm B}<-23$
and $0.3<z<2.3$ together with the prediction from the same model as in
the upper panel.  Here the model prediction has been corrected to take
into account the survey incompleteness.}
\label{fig:nmnzfit}
\end{figure}

We also fit to this data set a two-power-law LF and pure luminosity
evolution models with polynomial evolution in $L^*$ such that
$L^*(z)\propto10^{k_1z+k_2z^2}$.  The best-fit model has
$\Phi^*=0.2\times10^{-15}\,$Mpc$^{-3}\,$mag$^{-1}$, $\alpha=3.28$,
$\beta=1.08$, $M_{\rm B}^*=-21.45$, $k_1=1.41$ and $k_2=-0.29$ (see
Paper I).  However, even this best-fit model is formally rejected at
high significance ($>99$ per cent confidence) by the 2QZ data.  A
comparison between 2QZ data set used in the LF anlaysis and the best
fit model is shown in Fig \ref{fig:nmnzfit}.  This illustrates that
while the overall fit of the model to the data set is good, the cause
of the disagreement is a sharp peak in the redshift distribution of
2QZ QSOs at $z\simeq1.5$ which cannot be matched by the evolution
model.  This is likely to be caused by systematic variations in our
ability determine a QSO redshift. $z\simeq1.5$ corresponds to the
redshift where the \civ~ emission line moves into the observable
wavelength range of 2dF.  This strong emission line will increase the
probability of obtaining a good quality identification and redshift.

The new measurement of the QSO LF demonstrates that studies such as
these are now limited by the systematic errors, rather than Poisson
errors from small numbers of objects.  When the completed catalogue is
released, we will provide independent measurements of these systematic
effects, which can be derived from measurements of the efficiency of
our automated classification code.  We note that a preliminary
estimate of completeness as a function of $\bj$ and redshift (used
here and in Paper I) will be available electronically as part of the
data release.

\section{Data access}

The 2QZ 10k catalogue and associated data products are available from
{\tt http://www.2dfquasar.org} along with basic tools to access and
interrogate the catalogue.  A CD-ROM containing the same data and tools
is also available.  Any authors using the 2QZ 10k catalogue should
include the following  acknowledgement:  `The 2dF QSO Redshift Survey
(2QZ) was compiled by the 2QZ survey team from observations made with
the 2-degree Field on the Anglo-Australian Telescope'.

\section*{Acknowledgements} 

The 2QZ was based on observations made with the Anglo-Australian
Telescope and the UK Schmidt Telescope.  NL was supported by a PPARC
studentship during the course of this work.

\vspace{1.0truecm}
This paper has been produced using the Blackwell Scientific Publications 
\TeX macros.

\end{document}